\documentclass{llncs}

\usepackage{lineno,hyperref}

\newfont{\toto}{msbm10 at 12 pt}

\newcommand{\propagate}{{\em propagate}}
\newcommand{\collide}{{\em collide}}
\newcommand{\AoS}{{\em AoS}}
\newcommand{\SoA}{{\em SoA}}
\newcommand{\CSoA}{{\em CSoA}}
\newcommand{\CAoSoA}{{\em CAoSoA}}

\newcommand{\comment}[1]{}

\usepackage{varwidth}

\newcommand{\bc}{\begin{center}}
\newcommand{\ec}{\end{center}}

\newcommand{\bd}{\begin{description}}
\newcommand{\ed}{\end{description}}

\newcommand{\bi}{\begin{itemize}}
\newcommand{\ei}{\end{itemize}}

\newcommand{\benu}{\begin{enumerate}}
\newcommand{\eenu}{\end{enumerate}}

\newcommand{\bq}{\begin{quote}}
\newcommand{\eq}{\end{quote}}

\newcommand{\be}{\begin{equation}}
\newcommand{\ee}{\end{equation}}
\newcommand{\bea}{\begin{eqnarray}}
\newcommand{\eea}{\end{eqnarray}}

\usepackage{graphicx}
\usepackage{xcolor}
\usepackage{color}
\usepackage{listings}
\usepackage{bm}
\usepackage{url}
\usepackage{booktabs}

\definecolor{g90}{gray}{.95}
\definecolor{CadetBlue}{rgb}{0.28,0.23,0.54}
\definecolor{OliveGreen}{cmyk}{0.64,0,0.95,0.40}
\definecolor{Brown}{cmyk}{0,0.81,1,0.60}

\begin{document}



\title{
Early experience on using Knights Landing processors for Lattice Boltzmann applications
}


\author{
  Enrico Calore\inst{1,2}             \and
  Alessandro Gabbana\inst{1,2}        \and
  Sebastiano Fabio Schifano\inst{1,2} \and
  Raffaele Tripiccione\inst{1,2}
}

\institute{
Universit\`a di Ferrara, Ferrara, ITALY\\
INFN Ferrara, ITALY
}

\date{}


\maketitle


\begin{abstract}
{\em Knights Landing} (KNL) is the second generation 
of Intel processors based on {\em Many Integrated Cores} (MIC) 
architecture targeting HPC application segment.
It delivers massive thread and data parallelism together with high-speed 
on-chip memory bandwidth in a standalone processor that can boot a off-the-shelf 
Linux operating system.
KNL provides more than 3 TFlops of computing power for double-precision 
computation, doubling to 6 TFlops for single-precision.
In this work we assess the performance of this new processor for Lattice 
Boltzmann codes widely used in computational fluid-dynamics.
We design and implement an OpenMP code, and evaluate the impact of 
several data memory layouts to meet the different computing 
requirements of distinct parts of the application, aiming to exploit 
a large fraction of available peak computing throughput.
We also perform a preliminary analysis of energy efficiency, evaluating  
the time-to-solution and average-power consumption for each memory layout, 
and make some comparison with other processors and accelerators.    

\end{abstract}


\section{Introduction}\label{sec:intro}

Hi-end processors, the building blocks of HPC computer systems, have 
seen a steady increase in the number of processing cores, with cores 
able to perform more and more operations per clock-cycle. 
This trend has been further pushed forward in accelerators, such as GPUs 
and {\em Many Integrated Cores (MIC)} processors, offering large computing 
power together with a significant computing efficiency, e.g. a high ratio 
of computing power per Watt.
However, the use of accelerated systems is not without problems. 
The link between host CPU and accelerator, usually 
based on PCIe interface, creates a data bottleneck that reduces the 
sustained performances of most applications. Reducing 
the impact of this bottleneck in heterogeneous systems requires complex 
implementations~\cite{tang16,ijhpca17} with a non negligible impact 
on development an maintenance efforts. 


The latest generation Intel MIC accelerator, the {\em Knights Landing} 
(KNL) Xeon-Phi processor, offers a way out of this problem: it is
a self-hosted system, running a standard Linux operating system, so it
can be used alone to assemble homogeneous clusters.


In this work we present an early assessment of the performance of the KNL 
processor, using as test-case a state-of-the-art Lattice Boltzmann (LB) code. 
This application is very interesting for benchmarking purposes, as its
two main critical compute-intensive kernels, \propagate{} and \collide{}, 
are respectively strongly memory bound and compute bound.
For regular applications like LB codes, task parallelism is easily done by 
assigning tiles of the physical lattice to different cores.
However, exploiting data-parallelism through vectorization requires 
additional care, and in particular a careful design of the data layout 
is critical to allow an efficient use of vector instructions. 
Our code uses OpenMP to manage task parallelism, and we 
experiment with different data-layouts trying to find a compromise 
between the conflicting requirements of the  \propagate{} and \collide{} 
kernels.
We then assess the impact of several layout choices in terms of 
computing and energy performances.

Recent works have studied the performances of KNL~\cite{rosales16,li16,rucci17} with 
several applications, but as far as we know none of these investigate 
the impact of data layouts on performance and energy efficiency.

The rest of this paper is organized as following: \autoref{sec:knl-architecture} gives 
a short overview of the KNL architecture, highlighting the main 
features relevant for this work; \autoref{sec:lb} briefly sketches 
an outline of the Lattice Boltzmann method, while \autoref{sec:datastructure} 
presents the various options for data-layout that we have studied;
\autoref{sec:results} analyzes our results and ends with some concluding remarks.


\section{Overview of Knights Landing Architecture}\label{sec:knl-architecture}

The {\em Knights Landing} (KNL) is the second generation of Intel processors 
based on the MIC architecture, and the first self-bootable processor in this family.
It has an array of 64, 68 or 72 cores and four high speed memory banks 
based on the {\em Multi-Channel DRAM} (MCDRAM) technology 
providing an aggregate bandwidth of more than $450$ GB/s~\cite{stream}; 
it also integrates 6 DDR4 channels supporting up to $384$ GB of memory 
with a peak raw bandwidth of $115.2$ GB/s.
Two processors form a tile and share an L2-cache of $1$~MB; tiles are connected 
by a 2D-mesh of rings and can be clustered in several NUMA configurations.
%
%
In this work we only consider the {\em Quadrant} cluster configuration in which 
tiles are divided in four quadrants, each directly connected to one MCDRAM bank.
This configurations is the recommended one to use the KNL as a symmetric 
multi-processor, as it reduces the latency of L2-cache misses, and the 
4 block of MCDRAM appears as contiguous block of addresses. 
For more details on clustering see~\cite{colfax-knl-numa}.
MCDRAM on a KNL can be configured at boot time in Flat, Cache or Hybrid mode.
The Flat mode defines the whole MCDRAM as addressable memory allowing 
explicit data allocation, whereas Cache mode uses the MCDRAM as 
a last-level cache between the L2-caches and the on-platform DDR4 
memory. In Hybrid mode, the MCDRAM is used partly as addressable memory 
and partly as cache.
For more details on memory configuration see~\cite{colfax-knl-mcdram}.
In this work we only consider Flat and Cache modes.
Parallelism is exploited at two levels on the KNL: 
{\em task parallelism} builds onto the large number of integrated cores, 
while {\em data parallelism} uses the AVX 512-bit vector (SIMD) 
instructions.
Each core has two out-of-order vector processing units (VPUs) 
and supports the execution of up to 4 threads.
The KNL has a peak theoretical performance of 6 TFlops in single precision 
and 3 TFlops in double precision. 
Typical thermal design power (TDP) is $215$~W including 
MCDRAM memories (but not the Omni-Path interface).
For more details on KNL architecture see~\cite{sodani16}.

%
\begin{figure*}[!t]
\centering
\begin{minipage}{0.25\textwidth}
\includegraphics[width=0.9\textwidth]{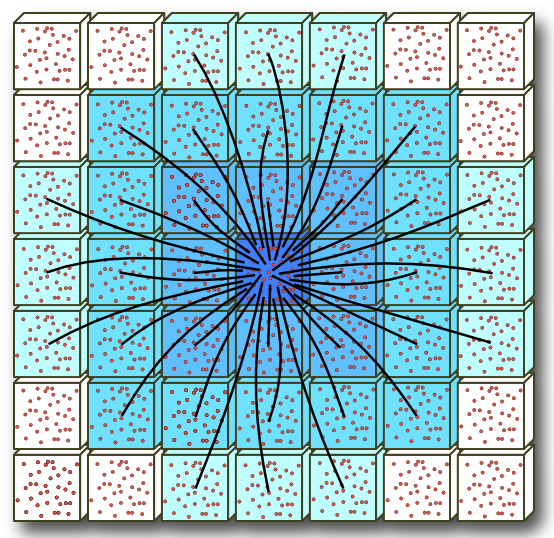}
\end{minipage}
\hspace*{20mm}
\begin{minipage}{0.25\textwidth}
\includegraphics[width=0.9\textwidth]{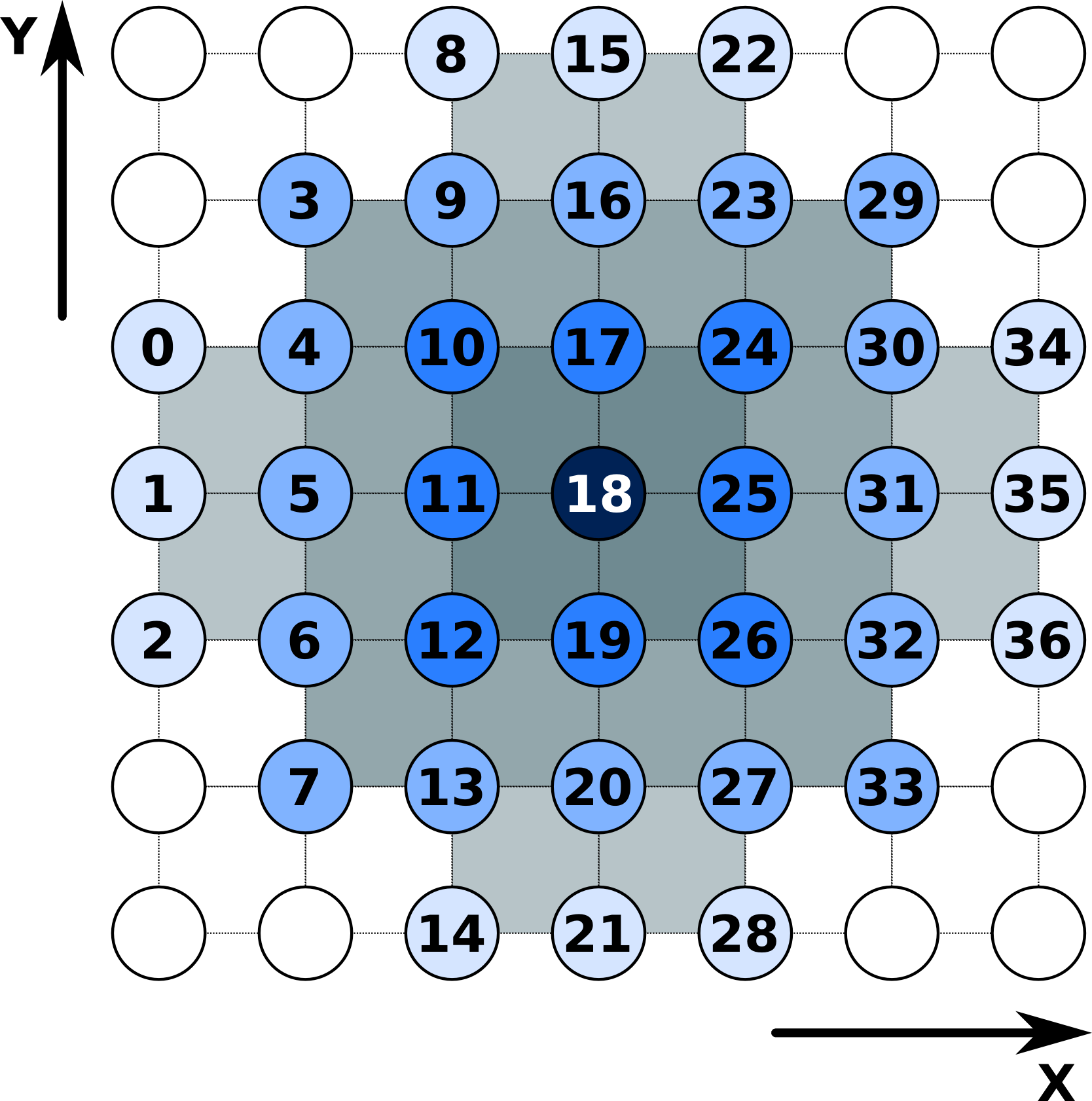}
\end{minipage}
\caption{Left: Velocity vectors for the LB populations in the D2Q37 model. 
Right: populations are identified by an arbitrary label, identifying the
lattice hop that they perform in the \propagate{} phase.}
\label{streamscheme}
\end{figure*}
%

\section{Lattice Boltzmann Methods}\label{sec:lb}

Lattice Boltzmann methods~\cite{sauro} (LB) are widely used in computational
fluid-dynamics, to describe fluid flows. LB methods are widely used in science
and engineering to accurately model single and multi-phase flows and 
can be easily accommodate irregular boundary conditions. This is why they are 
usually used in the oil\&gas industry to to study the 
dynamics of oil and shale-gas reservoirs and to maximize their yield. 

This class of applications, 
discrete in time and momenta and living on a discrete and regular grid of 
points, offers a large amount of available parallelism, so they are an ideal 
target for multi- and many-core processors. 
They are based on the synthetic dynamics of {\em populations} sitting at the 
sites of a discrete lattice. At each time step, populations \propagate{} 
from lattice-sites to lattice-sites, and then \collide{} mixing and 
changing their values accordingly.
In these processes, there is no data dependency between different lattice 
points, so both the \propagate{} and \collide{} steps can be performed in 
parallel on all grid points following any convenient schedule.

A model describing flows in $x$ dimensions and using $y$ populations is
labeled as $DxQy$.
In this work we study a D2Q37 model, a 2-dimensional system with 37 population 
associated to each lattice-site, corresponding to (pseudo-)particles moving up 
to three lattice points away, as shown in figure~\ref{streamscheme}.
This recently developed~\cite{JFM,POF} LB model automatically enforces 
the equation of state of a perfect
gas ($p = \rho T$); it has been recently used 
to perform large scale simulations of convective turbulence in several physics 
regimes~\cite{noi1,noi2}. 
The D2Q37 model is computationally more demanding than earlier methods; 
indeed, {\tt propagate} implies accessing 37 neighbor cells to gather 
all populations, while {\tt collide } executes $\approx 6600$ 
double-precision floating point operations per lattice point.

%
\begin{figure}[t]
\center
\includegraphics[width=\textwidth]{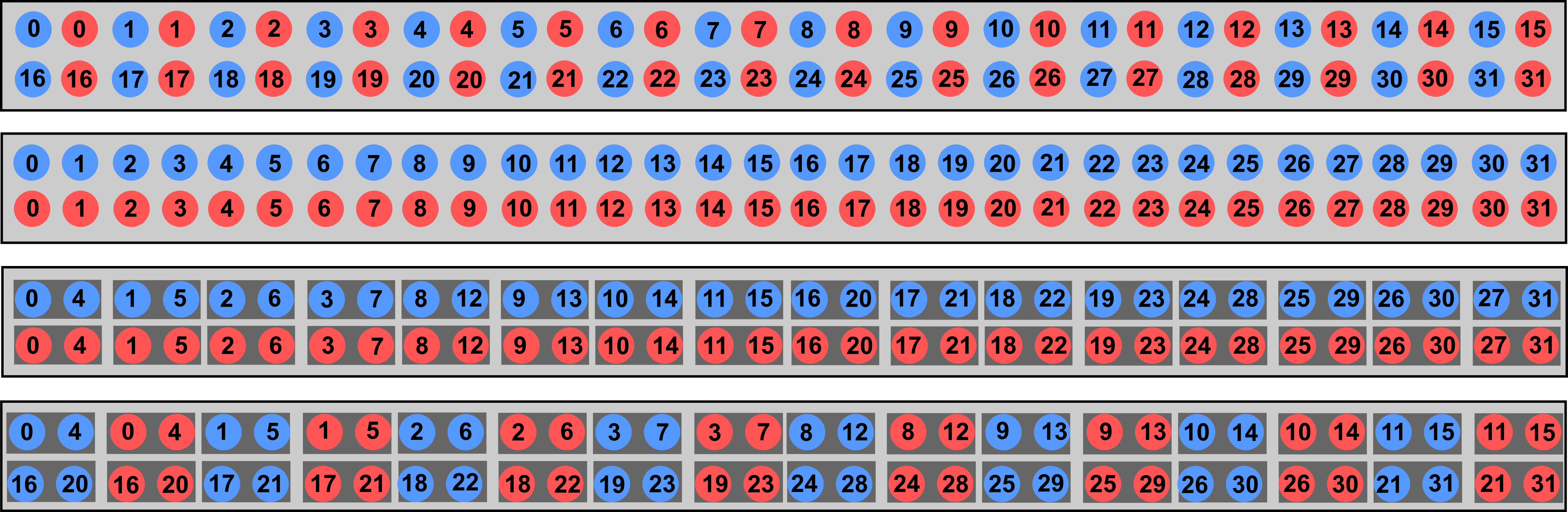}
\caption{
Top to bottom, \AoS{}, \SoA{}, \CSoA{} and \CAoSoA{} data memory layouts 
for a $2\times8$ lattice with two populations (red and blue) per site 
and VL=2. For \CSoA{} and \CAoSoA{} 
each grey-box is a cluster. Memory addresses 
increase left-to-right top-to-bottom. 
\label{fig:data-structures}
}
\end{figure}
%


\section{Implementation and Optimization of D2Q37 LB model}\label{sec:datastructure}

In all LB methods \propagate{} and \collide{} use most of compute 
cycles of the whole application so optimization efforts have to focus 
largely on these two kernels. 
Data allocation policies and memory-layout decisions are becoming more 
and more important for performance on recent many-core processors; this 
is even more so, as we have to find a data layout that matches the conflicting 
requirements of these two kernels.
In this section we focus mostly on this point, discussing several possible 
choices and showing that they have very large effects on the obtained 
performance (and energy-performance) for the KNL processor; here, we 
extend previous works~\cite{ppam15,ijhpca17}, where additional details 
on other aspects of the code structure are available.

{\em Array of Structures} (\AoS{}) or {\em Structure of Arrays} (\SoA{}) 
offer a starting point to consider more complex data memory organizations.
In the \AoS{} scheme, population data associated to each lattice site 
are stored one after the other at contiguous memory addresses. 
In this arrangement all data associated to one lattice point are at close
memory locations, but same index populations of different lattice sites are
stored in memory at non-unit strided  addresses; this makes it more difficult 
to process them using vector SIMD instructions.
Conversely, the \SoA{} scheme stores same index populations of all sites one
after the other; this is appropriate for vector SIMD instructions, as it allows
to move several lattice sites -- $8$ for the KNL -- in parallel. 
\autoref{fig:data-structures} -- first two designs at the top -- visualize the
\AoS{} and \SoA{} layouts, for a sample case of a lattice of $2 \times 8$ 
sites with only two populations (red and blue). 
%
%
\begin{figure}[t]
\centering
%
%
\begin{lstlisting}[basicstyle=\scriptsize,language=C]
#define LYOVL (LY / VL)
typedef struct { double c[VL]; } vdata_t;
typedef struct { vdata_t s[LX*LYOVL]; } vpop_csoa_t;
vpop_csoa_t prv[NPOP], nxt[NPOP];
#pragma omp parallel for num_threads(NTHREAD) schedule(dynamic)
for ( ix = startX; ix < endX; ix++ ) {
  idx = (NYOVL*ix) + HYOVL; 
  for( p = 0; p < NPOP; p++){
    for ( iy = 0; iy < SIZEYOVL; iy++ ) {      
      #pragma unroll
      #pragma vector aligned nontemporal
      for(k = 0; k < VL;k++) 
      	nxt->p[p][idx+iy].c[k] = prv->p[p][idx+iy+OFF[p]].c[k]
} } }
\end{lstlisting}
\caption{
\label{code:propagate-csoa}
Source code of {\tt propagate} kernel for using the \CSoA{} data layouts.
{\tt OFF} is a vector containing memory-address offsets associated 
to each population hop. {\tt VL} is the size of a cluster. 
}
\end{figure}
%
%
%
%
This layout has a potential inefficiency associated to 
unaligned memory accesses; in fact, the read-address for  
population values is computed as the sum of the address of the current site 
plus an offset, and the resulting address is in general not aligned to 
a $64$ Byte boundary, preventing direct memory copies to vector registers.

In order to circumvent this problem, we start from the \SoA{} layout and, for a
lattice of size $LX \times LY$, we cluster together {\tt VL} elements of each
population at a distance $LY/VL$, with {\tt VL} a multiple of the KNL vector
size. 
We call this data layout a {\em Cluster Structure of Array} (CSoA), 
see \autoref{fig:data-structures} -- third design from top -- for the case of
$VL=2$  corresponding to an hypothetical processor using vectors of length $2$. 
Using \CSoA{}, \propagate{}, whose main task is to read the same population 
elements at all sites and move them to different sites, is able to use vector 
instructions to process clusters of properly memory-aligned items.
\autoref{code:propagate-csoa} shows the corresponding {\tt C} 
type definitions and code implementation for \propagate{}. 
%
%
The loop on X is parallelized among the threads using the OpenMP pragma
parallel loop, making each thread to work on a slice of the lattice;  the inner
loop, copying elements of a cluster into another cluster, can be unrolled and
vectorized since both read and write pointers are now properly aligned. 
A further optimization can in this case be applied with the use of non-temporal 
write operations saving time and reducing the overall memory traffic 
by $1/3$~\cite{ijhpca17}.
We instruct the compiler to use these optimizations using pragmas {\tt unroll} 
and ({\tt vector aligned nontemporal}).
%
%
%
%
\begin{figure}[t]
\includegraphics[width=\textwidth]{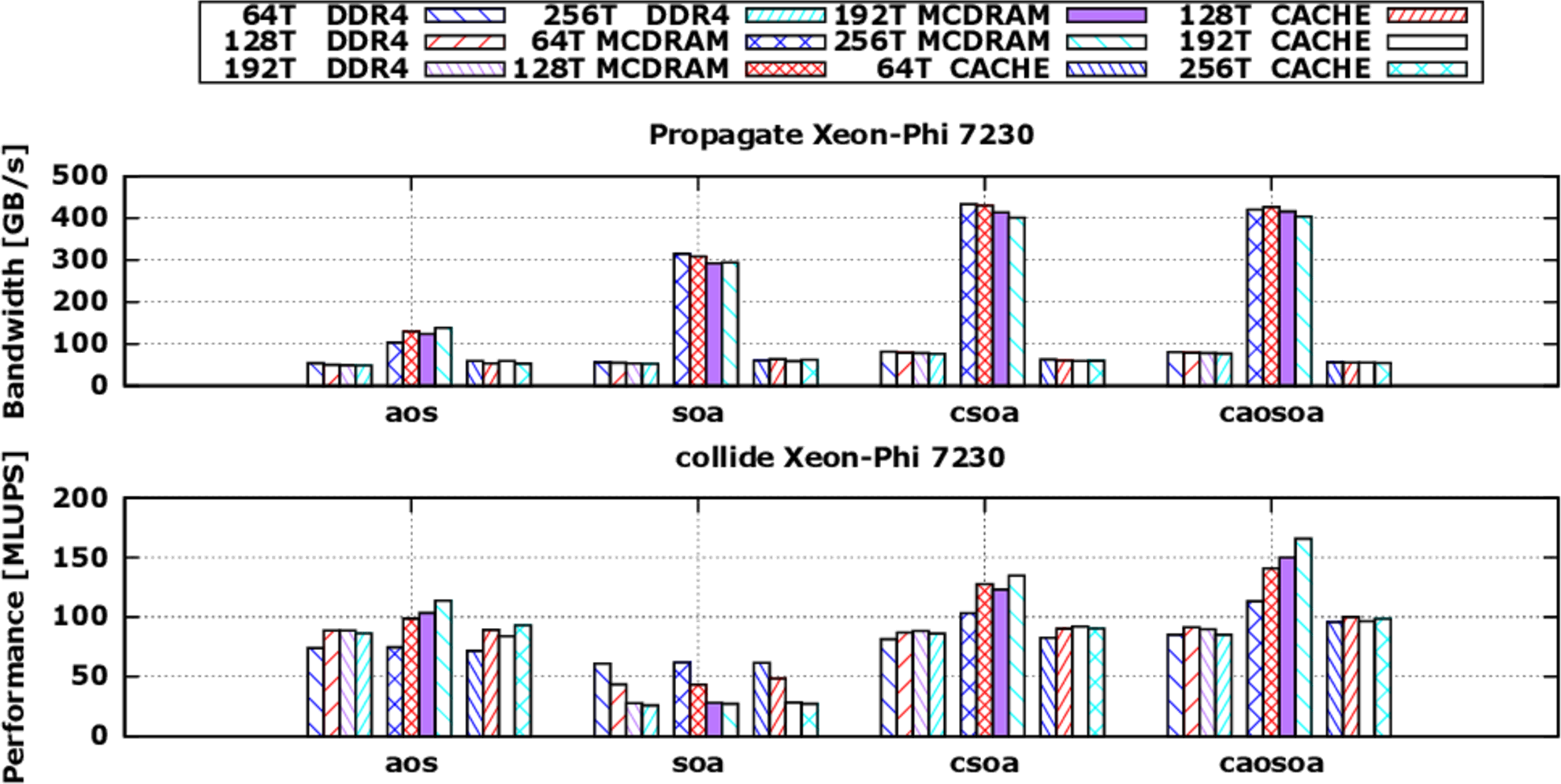}
\caption{
\label{fig:performance}
Performance of \propagate{} (top) and \collide{} (bottom) using 
the \AoS{}, \SoA{}, \CSoA{} and \CAoSoA{} data layouts. Performance 
for \propagate{} is shown in MLUPS, defined in the text.
All data for a 64 core Xeon-Phi 7230 running at $1.4$ GHz.
For the FLAT configuration we use a $2304\times 8192$ lattice  
that fits into MCDRAM; for the CACHE configuration, the lattice 
is $4608\times 12288$, twice the size of MCDRAM. For each layout, 
3 groups of 4 bars correspond respectively to FLAT-DDR4, FLAT-MCDRAM 
and CACHE. 
Within each group, bars correspond respectively to 1,2,3 and 4 
threads per core. 
}
\end{figure}
%
%
\autoref{fig:performance} shows measured bandwidth for our data structures,
using the FLAT memory mode, and using both off-chip or MCDRAM memory, and the
CACHE memory mode. Data refer to a 64 core Xeon-Phi 7230 running at $1.4$ GHz. 

The \collide{} kernel can be vectorized using the same strategy 
as of \propagate{}, so one expects the \CSoA{} layout should be 
an efficient choice; however, profiling the execution of this kernel, 
we found that a large number of TLB misses are generated. 
These happens because different populations associated to each lattice site 
are stored at memory addresses far from each other, and several non-unit stride 
reads are necessary to load all population values necessary to compute 
the collisional operator.
We then introduce yet another data layout, in order to reduce this penalty.
We start again from the \SoA{} layout, and for each population array, 
we divide each $Y$-column in {\tt VL} partitions each of size {\tt LY/VL}; 
all elements sitting at the $i$th position of each partition are then packed 
together into an array of {\tt VL} elements called {\em cluster}. 
For each index $i$ we then store in memory one after the other the 
37 clusters -- one for each population -- associated to it. 
%
%
%
%
This defines the {\em Clustered Array of Structure of Arrays} (CAoSoA); 
the main improvement on \CSoA{} is that it still allows vectorization 
of clusters of size $VL$, and at the same time improves locality of 
populations, keeping all population data associated to each lattice site at close 
and aligned addresses; see again \autoref{fig:data-structures} for a visual
description.
%
%
%
%
This data layout combines the benefits of the \CSoA{} scheme, allowing 
aligned memory accesses and vectorization (relevant for the \propagate{} 
kernel) and at the same providing population locality (together relevant 
for the \collide{} kernel).
%
%
%
%
%
%
%
\autoref{fig:performance} shows measured performances for the 
\collide{} kernel -- expressed in {\em Million Lattice UPdates per Second}, 
a common figure of performance for these codes  -- for all data-layouts 
considered sofar.
For a lattice of $2560 \times 8192$ sites, using  \CAoSoA{} we have 
reduced to zero the number of TLB misses of \collide{} measured using the 
hardware counter {\tt PAGE\_WALKS.D\_SIDE\_WALKS}) w.r.t. almost 
$2$ billions misses for \CSoA{}; correspondingly, the number of clocks 
ticks (counter {\tt CPU\_CLK\_UNHALTED.REF\_TSC}) decreases by 
approximately 25\%.  
The picture also shows that the performance of the \propagate{} kernel 
is unchanged using the \CSoA{} and \CAoSoA{} layouts.


\section{Analysis of Results and Conclusions}\label{sec:results}

%
%
We start summarizing our performance data. See again \autoref{fig:performance} 
showing results for \propagate{} and \collide{}, using the FLAT 
and CACHE memory configurations.
%
%
%
For the \propagate{} kernel, performance is almost independent 
from the number of threads per core, while the impact of the 
various data layouts is large; indeed, using a FLAT MCDRAM 
configuration the measured bandwidth increases from $138$ GB/s 
of \AoS{} to $314$ GB/s of \SoA{} and to $433$ GB/s  of \CSoA{}. 
This trend is similar using the DDR4 memory banks but performance 
is much lower, ranging from $54$ GB/s of \AoS{} to $56$ GB/s 
of \SoA{} and to $81$ GB/s of \CSoA{}.  
We have a similar behavior also with the CACHE configuration, measuring  
in this case a bandwidth of $59$, $60$ and $62$ GB/s for the 
\AoS{}, \SoA{} and \CSoA{} memory layouts for a lattice size 
that does not fit into MCDRAM.
Using the \CAoSoA{} layout, performance does not further improves, 
both for FLAT and CACHE configurations. 

%

For \collide{} kernel, using a FLAT MCDRAM configuration 
we obtain a good level of performance, $114$ MLUPS,  using the \AoS{} 
layout with 4 threads per core; the \SoA{} layout performance does not 
allow efficient vectorization, so performance goes down to $62$ MLUPS 
with one thread per core, further decreasing if we use 2, 3 and 4 
threads per core. 
Enforcing memory alignment with the \CSoA{} layout, we obtain again a 
properly vectorized code  and performance increases up to $135$ MLUPS 
using 4 threads per core. 
Performances 
further improve with the \CAoSoA{} layout as we remove the overhead 
associated to TLB misses and we reach the level of $165$ MLUPS with 4 
threads per core, 
corresponding to a factor $1.4$X and $1.2$X w.r.t. the \AoS{} and \CSoA{} 
layouts.
As the \collide{} kernels performs approximately $6600$ floating-point 
operations per lattice size, our KNL processor, using the \CAoSoA{} layout, 
delivers a sustained performance of approximately $1$ TFlops, that is 
about $30\%$ of the available raw peak.
If one used DDR4 memory performances are harmed by memory bandwidth, 
but results follows the same trend as in the MCDRAM case, reaching 
$89$ MLUPS with the \CAoSoA{} layout. 
The same is true with the CACHE configuration where 
\collide{} reaches a peak of $98$ MLUPS for the \CAoSoA{} layout.

%
\begin{figure}[t]
%
%
\includegraphics[width=\textwidth]{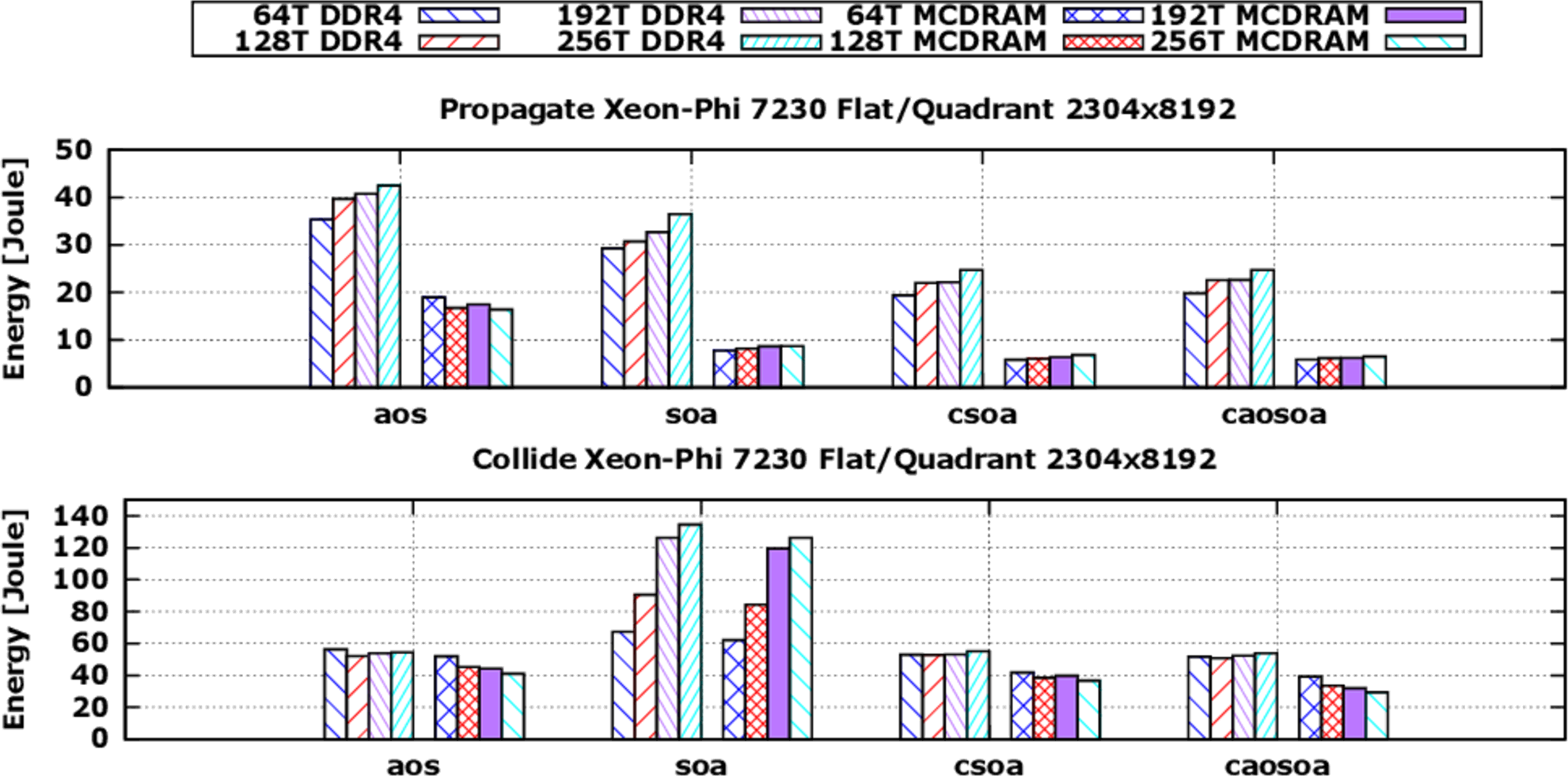}
\caption{
\label{fig:energycompare}
Energy-to-Solution for \propagate{} (top) and \collide{} (bottom), 
for all data layouts, using the FLAT configuration. 
For each layout we plot two groups of bars corresponding to the use of 
either DDR4 off-chip memory or on-chip MCDRAM. Within each group the bars 
correspond respectively to 1,2,3 and 4 threads per core. 
All values are computed as the sum of the \textit{Package} and \textit{DRAM} 
RAPL energy counters, per iteration.
}
\end{figure}
%
%

We now consider energy efficiency for our code; we use data from
the RAPL (Running Average Power Limit) registers 
available in the KNL processor, for both \textit{Package} and \textit{DRAM}
counters, that we read using the custom library described in~\cite{ccpe17}.
Results are shown in \autoref{fig:energycompare} for both FLAT memory
configurations,  highlighting the impact of  data-layouts on energy consumption.
All figures refer to {\em Energy-to-Solution} ($E_S$) and are the 
sum of \textit{Package} (in-chip) and \textit{DRAM} (off-chip memory) 
contributions.
For \propagate{}, we see that using MCDRAM increases the average 
power drain ($\approx 35\%$) compared to the use of off-chip DDR4, 
but $E_S$ is lower  since a slightly higher power gets 
integrated over a much shorter ($\approx 4 \times$) time. 
Also, the \CSoA{} and \CAoSoA{} data-layouts halve $E_S$ 
w.r.t. the \AoS{} and \SoA{} layouts as a result of their 
shorter execution times and slightly lower power drain.
For the \collide{} kernel the \SoA{} layout has a rather low power drain 
($\approx 30\%$ less than \CSoA{} and \CAoSoA{}) because vector units 
are not used; however, the code runs also much slower ($\approx 3 \times$),
translating into the worst performance figure in terms of $E_S$. 
Conversely, the \CAoSoA{} layout gives the best result in term of
energy efficiency, with energy-to-solution decreasing while increasing the 
number of threads per core, thanks to a constant power drain and an increasing
performance.
Using CACHE configurations, the average power drain is in between the values 
recorded for the DDR4 and MCDRAM cases.
As shown in \autoref{fig:performance} performances are similar to the case of 
DDR4, with a slightly performance decrease for \propagate{} and a slightly
increase for \collide{} when using \CSoA{} and \CAoSoA{} data-layouts.
Thus, from the energy consumption point of view, using cache configuration 
leads to similar energy behaviors as using DDR4.

%
\begin{table}[t]
\caption{
\label{tab:comparison}
Performance comparison among several processors. We consider the \propagate{} 
and \collide{} kernels and the full code (Global), using
the \CAoSoA{} data layout. We compare the KNL against the MIC KNC, 
the NVIDIA GK210 and P100 GPUs, and the Intel E52697v4 CPU. 
The row labeled with {\em Global} report the perforamnce of the full code. 
}
\centering
\resizebox{\textwidth}{!}{
\begin{tabular}{lrrrrrrrrrrrrrr|rr}
\toprule
                        && KNC 7120P && GK210 && P100 && E52697v4 && KNL 7230  && KNL 7230   &&& KNL 7230   \\
                        &&	     &&       &&      &&          && flat/quad && cache/quad &&& cache/quad \\
\midrule
Lattice size            && \multicolumn{11}{c}{$1024\times 8192$}                            &&& $4608 \times 12288$ \\
Memory footprint [GB]   && \multicolumn{11}{c}{$\approx 4.6$}                                &&& $\approx 30$ \\
\midrule
$T_{\mbox{prop}}$ [ms]  && 49.9      && 32.3  && 12.5 && 98.06    &&  12.5     && 19.65      &&& 506.64     \\
$T_{\mbox{coll}}$ [ms]  && 180.9     && 71.1  && 24.1 && 173.42   &&  50.3     && 51.42      &&& 550.25     \\
\midrule
Propagate [GB/s]        && 100       && 155   && 396  && 51	  &&  398      && 253	     &&& 66	    \\
Collide [GF/s]          && 307       && 764   && 2253 && 320	  &&  1100     && 1079       &&& 680	    \\
Collide [MLUPS]         && 46	     && 115   && 340  && 48	  &&  166      && 163	     &&& 103	    \\
\midrule
Global [MLUPS]          && 35	     &&  73   && 232  && 31	  &&  119      && 106	     &&& 67	    \\
\bottomrule
\end{tabular}
}
\end{table}
%

We finally compare our performance results with that of other recent 
multi- and many-core processors~\cite{caf13,cafgpu13,parco16}.
Our comparison is shown in \autoref{tab:comparison} for both critical 
kernels and also for the complete code; we adopt the \CAoSoA{} layout 
throughout, as it offers the best performance. 
Let first discuss the case of lattice size $1024\times 8192$ requiring  
a memory footprint of $\approx 4.6$ GB fitting the  $16$ GB on-chip 
MCDRAM. The data size also fits most other accelerator boards, 
so we can perform a meaningful comparison. 
%
%
%
Comparing the KNL in FLAT mode with the KNC 7120P, the previous 
generation MIC processor, we see that performances for \propagate{} 
and \collide{}  are respectively $\approx 4$X and $\approx 3.5$X faster.
Comparing with NVIDIA GPUs, the execution time for \propagate{} is 
$\approx 2.5$X faster than on a GK2010 GPU (hosted on a K80 board), 
and the same as a P100 Pascal board. 
The execution time of \collide{} is $1.4$X faster than a GK210, 
and approximately $50$\% slower than a P100.
Comparing performances with a more traditional Intel E5-2697v4 CPU, 
based on Broadwell  micro-architecture, \propagate{} is $7.8$X faster 
and \collide{} is $3.5$X faster.
Using the KNL in CACHE mode with a lattice that does not fit into MCDRAM, 
the performance of the processor are much slower. In the last column 
at right of \autoref{tab:comparison} we see the results for a lattice 
using a memory footprint twice the size of MCDRAM. 
In this case, comparing with CPU E5-2697v4 for which the lattice 
$1024\times 8192$ does not fit in the last-level cache, performances  
of \propagate{} are more or less the same, and that of \collide{} 
are $\approx 2$X faster. 

In summary, based on our experience related to our application, 
some concluding remarks are in order:
i) the KNL architecture makes it easy to port and run codes previously 
developed for X86 standard CPUs; however performance is strongly 
affected by the massive level of parallelism that must necessarily 
be exploited on the processor, lest the level of performance drops 
to the value of standard multi-core CPUs or even worst;
ii) for this reason data layouts plays a relevant role in allowing 
to use an efficient level of vectorization; at least for LB applications, 
appropriate data structures are necessary to allow the different 
vectorization strategies necessary in different parts of the application;  
iii) the KNL processor improves on the KNC -- the previous generation 
MIC processors -- by a factor $\approx 3-4$X; 
iv) if application data fits within the MCDRAM, 
performances are very competitive with that of GPU accelerators; 
however, if this is not the case, performance drops to levels 
similar to those of multi-core CPUs, with the further drawback 
that codes and operations (editing, compilations, IO, etc.) 
not exploiting task and data parallelism run much slower.

In the future, we plan to further analyze the energy performances of KNL 
comparing with other processors, and to design and develop a parallel 
hybrid MPI+OpenMP code able to run on a cluster of KNLs, in order to 
investigate scalability.


{\small
\subsubsection*{Acknowledgements.}
This work was done in the framework of the COKA, COSA and SUMA projects 
of INFN. We would like to thank CINECA (Italy) for access to their 
HPC systems. AG has been supported by the EU Horizon 2020 research and
innovation programme under the Marie Sklodowska-Curie grant agreement 
No 642069.
}





\end{document}